# A Toolkit for Compliance, a Toolkit for Justice: Drawing on Cross-sectoral Expertise to Develop a Pro-justice EU AI Act Toolkit

Tomasz Hollanek*
University of Cambridge
Cambridge, United Kingdom
th536@cam.ac.uk

Yulu Pi
University of Warwick
Warwick, United Kingdom
Yulu.Pi@warwick.ac.uk

Cosimo Fiorini
Accenture
Modena, Italy
cosimo.fiorini@accenture.com

Virginia Vignali
Accenture
Modena, Italy
virginia.vignali@accenture.com

Dorian Peters
Imperial College London
London, United Kingdom
d.peters@imperial.ac.uk

Eleanor Drage
University of Cambridge
Cambridge, United Kingdom
ed575@cam.ac.uk

## Abstract

The introduction of the AI Act in the European Union presents the AI research and practice community with a set of new challenges related to compliance. While it is certain that AI practitioners will require additional guidance and tools to meet these requirements, previous research on toolkits that aim to translate the 'theory' of AI ethics into development and deployment 'practice' suggests that such resources suffer from multiple limitations. These limitations stem, in part, from the fact that the toolkits are either produced by industry-based teams (often presenting 'ethical practice' as a box-ticking compliance exercise) or by academics whose work tends to be abstract and divorced from the realities of industry. In this paper, we discuss the challenge of developing an AI ethics toolkit for practitioners that helps them comply with new AI-focused regulation, but that also moves beyond mere compliance to consider broader socio-ethical questions throughout development and deployment. The toolkit was created through a cross-sectoral collaboration between an academic team based in the UK and an industry team in Italy. We outline the background and rationale for creating a pro-justice AI Act compliance toolkit, detail the process undertaken to develop it, and describe the collaboration and negotiation efforts that shaped its creation. We aim for the described process to serve as a blueprint for other teams navigating the challenges of academia-industry partnerships and aspiring to produce usable and meaningful AI ethics resources.

## 1 INTRODUCTION

The introduction of the AI Act in the European Union presents the AI research and practice community with a set of new challenges. For instance, under the Act, AI providers must decide whether their system poses risks of a prohibited, high, or limited nature based on a classification system that distinguishes AI systems based on their capabilities and intended application areas. Those who self-identify as providers and deployers of 'high-risk AI' must then adopt a number of special precautions aiming to ensure that the risks stemming from a given system's development and deployment are mitigated.

While it is certain that AI practitioners will require additional guidance to meet these requirements, previous research on tools that aim to translate the *theory* of AI ethics (including high-level principles and values) into development *practice* suggests that they suffer from multiple limitations and are rarely, if ever, used in industry [15, 25, 30, 32]. These limitations, which we discuss further in the following section, stem in part from the fact that toolkits and guidelines on responsible AI are most often generated by either academics or by industry-based teams, but rarely by both in collaboration. Siloed approaches risk missing the mark by inadvertently neglecting the expertise of all audiences. In this paper, we describe and analyse the process we followed to co-create a comprehensive AI ethics toolkit in a partnership between industry-based practitioners and academic researchers specialising in human-computer interaction, feminist ethics, and critical design. The partnership approach allowed for the preservation of ethical rigor while responding to the practical needs of industry development.

Throughout a year-long period, a team of UK-based academic researchers from across disciplines (henceforth, 'AT' for Academic

Team) joint together with a team of technology developers, designers, and product managers from an AI technology company based in Italy (henceforth, 'IT' for Industry Team). The objectives of the project were twofold. Firstly, we wanted to co-create a toolkit that would help developers of high-risk AI (especially from small to medium organisations) comply with the AI Act. From the start, both teams were keen to take a pro-justice approach that would offer enhanced support (above and beyond compliance) for issues like disability justice and redress which remain merely voluntary or vague within the Act. The theoretical foundations for this pro-justice approach are summarised in the next section and elaborated in greater detail in a separate article [17]. Secondly, our aim was to develop and test a method for navigating a balanced partnership between academics and industry professionals – a partnership that ran the full length of the project and which brought a combination of technical and socio-technical expertise to bear on the AI development process. The AT and IT held weekly online meetings to address misalignments and roadblocks, supplemented by two multi-day in-person co-design sessions. The IT not only developed the toolkit application but also actively contributed to the development of the pro-justice content by contributing case studies and examples from their practical experience as an AI development team. We also engaged with a large advisory board of interdisciplinary experts who provided written feedback in highly specialised areas (including privacy law, machine learning, and disability and design) to guide and course-correct as the project progressed.

In this paper, we outline the background and rationale for creating a pro-justice compliance tool, detail the process undertaken to develop it, and describe the collaboration and negotiation efforts that shaped its creation. We aim for the described process to serve as a blueprint for other teams navigating the challenges of academia-industry partnerships and/or aiming to develop effective AI ethics tools. To this end, we reflect on the challenges and tensions, and present insights and lessons learned along the way that may shed light, not only on one method for developing responsible AI tools, but on why deep cross-sectorial teamwork is important for complex research-to-practice translational work.

## 2 BACKGROUND AND RATIONALE

### 2.1 Our pro-justice approach to AI ethics and policy

Both the AT and the IT involved in this project were independently invested in applying diverse pro-justice ethical frameworks to AI ethics and policy considerations. The AT specialise in using anti-racist and feminist theory and praxis to explore technology's impact on, and entanglement with, preexisting social, economic, and environmental conditions. The IT believes in a human-centric approach to technology, that must be accessible and interpretable by everyone, and views AI as a tool for informed decision-making, designed to enhance human capabilities without causing harm to its users. Both the AT and the IT believe that feminism and other social movements that address social inequality should be foundational to the development and deployment of AI, because only in shifting the balance of power can AI have a positive effect. Feminism has played a central role in holding those who create and deploy AI accountable for unjust and harmful outcomes, and in offering different avenues for innovation [7] (pp. xii, xxi). We explain this approach further and elaborate on how it affected our interpretation of the AI Act – in a way that animates the spirit as well as the letter of the law – in a separate paper [17]. In short, we defined obligations set out by the EU through feminist ideas and practices, which involved race, gender, class, disability, age and environmental justice-sensitive approaches to design. These were selected as an antidote to 'apolitical', vague, and ambiguous AI ethics principles that 'create real confusion among tech workers' who interpret them in many different ways, making them 'unable to do the ethical work they are intended to do' [8].

In this paper, we deal more directly with how this conception of ethico-politics in technology development – prioritised by the AT involved in this project – informed the co-design process of the whole team, including industry partners. It is because of this commitment to pro-justice approaches that the following questions were at the core of the entire process: Can the EU AI Act be feminist? How can the EU AI Act – which has been accused of having been 'watered down' by consistent corporate lobbying – be reinvigorated? How can this ethical approach be made meaningful to and actionable for practitioners?

### 2.2 Our approach to toolkits as instruments of ethics and compliance

While one of the direct objectives of this project was to co-create an AI ethics toolkit for AI Act compliance, we should also note that our process was informed by critiques of what we could refer to as 'toolkitification' of design methods in general and ethical AI in particular [3, 16, 20, 23]. Toolkitification refers to the proliferation of various types of resources – 'toolkits', 'guidelines,' or 'blueprints,' terms used largely interchangeably – ranging from the technical and procedural to the educational and meant to help translate AI ethics theory into design and development practice. Prior to the start of the project, one of this paper's authors conducted a study of the 'landscape' of AI ethics toolkits to understand what resources were already available, what their limitations were, and how the toolkit format could be applied effectively to AI ethics and compliance.

Previous studies of the AI ethics toolkits landscape have explored the alignment of values, principles, and goals across institutions, sectors, and regions [4, 11, 19]. Some focus on toolkit usability [21, 29] or compare the formats and methods adopted [5, 24]. Other research highlights the ethical traditions the toolkits draw from, the stage at which 'ethics work' should be carried out, and the stakeholders involved [4, 13]. A meta-analysis of these various studies found several consistent issues with the toolkit format, including: gaps in the lists of values that toolkits are supposed to help translate into practice (i.e. 'solidarity' and 'environmental sustainability'); a simplistic view of ethics embodied by the toolkit format; and the co-optation of terms like *participatory design* in ways that hinder meaningful adoption [15]. To address these limitations, researchers suggest that future AI ethics toolkits should:

1. Incorporate diverse ethical perspectives, such as feminist and Indigenous viewpoints.
2. Position ethics as a collective process involving both design and business decisions.

3. Scaffold ethical deliberation while acknowledging its complexity.
4. Reject box-ticking approaches and embrace power-sensitive, participatory design practices [15, 32].

These suggestions informed our own idea for the role of the toolkit in ensuring responsible, EU AI Act-compliant AI. The toolkit in this context is a means of providing scaffolding for the process of ethical deliberation in technology design and making compliance more easily executable – but without making the process seem simple or free from friction [16].

### 2.3 Our approach to academia-industry collaboration

Research into AI ethics toolkits and guidelines indicates that they are rarely used by industry professionals, in part because many toolkits are created in academic environments without input from industry experts, leading to a disconnect with practitioners' needs [25, 30] Further, the experience from industry that informs our project also highlights that the checklist format – which many AI ethics tools/guidelines adopt [14] – is often difficult to interpret, designed by compliance experts in a technical language not interpretable by other business units, and still lengthy and tedious to implement. The result is the bypassing or underestimation of important decisions in the design/development phase.

This project emerges from a deliberate collaboration between the AT and IT precisely to avoid such common pitfalls. Our approach to the academia-industry partnership was guided by the following principles:

- **Value alignment**: Establishing a shared vision for the toolkit from the outset to ensure both teams were aligned on the project's goals and values.
- **Consistent collaboration and timely adjustments**: Ensuring ongoing engagement and communication throughout the project, with a focus on making adjustments as needed based on feedback and evolving needs.
- **Grounding practical application in sound theoretical foundations**: Ensuring that the toolkit was not only practical and user-friendly but also deeply informed by ethical frameworks and academic insights.
- **Holistic expertise**: Engaging a diverse range of disciplinary perspectives to inform the development of the toolkit, drawing on both academic and industry expertise.

While the value of an industry-academia partnership is clear, in light of the issues mentioned above, such collaborations also come with a number of challenges, related in particular to a potential mismatch between the two teams' incentives, constraints and work cultures. To ensure transparency and enable critical reflection on both the nature of the collaboration and the emerging results, it is important to account for how the partnership was established and what both teams agreed upon in initial discussions, in particular what values and principles were to guide the collaboration moving forward.

The first contact was made by representatives from the research unit of an Italian AI provider (the IT), who reached out in late 2022 to the academic research centre in the UK where the AT team is based. The IT's interest was in forming a potential collaboration to develop an AI Act compliance tool. It is important to note that the IT was already well-acquainted with the work of the AT, which extends beyond legal and technical considerations to deeply engaging with the social and ethical dimensions of AI development and deployment. Furthermore, while the IT is part of a commercial company, it had previously engaged in academic and civil society projects and had received funding from the Emilia-Romagna region to support the development of an open access AI Act compliance tool.

At the time of initial contact, the AT was already working conceptually on a tool to support compliance with the EU AI Act. As such, the team was open to exploring a collaboration with industry partners. A key consideration for the AT, however, was that the tool, while practical and user-friendly, should integrate pro-justice ethical frameworks into the toolkit's foundation. This commitment to a pro-justice lens on both sides established value alignment between the two teams from the outset and has been crucial in shaping the collaboration as a process of mutual learning. Early discussions established mechanisms to guide the partnership, such as regular meetings and open communication to ensure continual alignment with the shared vision.

During the course of the project, the IT was acquired by a global consulting firm and underwent a leadership transition. Despite this shift, their commitment to continuous collaboration allowed both teams to make necessary adjustments and realign their values as the partnership evolved. Details of the negotiation and decision-making process are outlined in the Process section below, as well as in section 5.3.1.

Additionally, while the core team (IT and AT) brings a diverse range of expertise critical to the project's success, we recognised that we needed to consult with experts from a broader range of disciplines and practices related to AI ethics, AI development, and the EU AI Act. To this end, we established an Advisory Board that would gesture towards the range of expertise that one needs to develop a holistic response to AI Act compliance. We invited representatives from both Big Tech and smaller AI companies to be inclusive and recognise that pro-justice work happens – albeit at varying levels of concentration – at all corners of the AI landscape. We curated our Board so as to include as broad a group as possible of experts with respect to their areas of specialism, and who could advise us on how to action our principles in a variety of ways. External advice at this stage is as crucial as independent auditing later in a product's lifecycle in disrupting the echo chamber that can emerge in close-knit teams. Despite being AI ethics experts ourselves, we cannot assume that a single team is enough: a system of checks and balances is essential. This is why AI ethics expertise is collective, rather than individual: a single consultant will not suffice because AI ethics knowledge emerges through interactions between people with different insights into how a system functions and impacts society.

## 3 PROCESS

### 3.1 Ongoing collaboration and co-design between academia and industry

The collaboration between the IT and the AT relied on consistent, structured engagement over the course of a year and half, with the full team meeting weekly online to align objectives, resolve issues, refine methods and review progress. This steady involvement was complemented by two sessions of two-day, in-person co-design activities that served as milestones in the toolkit's development.

### 3.1.1 Value elicitation and alignment; user proto-persona development; toolkit format-scoping.
After an initial negotiation of collaboration terms and intellectual property arrangements, the two teams convened for an in-person workshop, which focused on three key elements:

•**Value elicitation and alignment**: This session allowed both teams to verbalise their objectives, explore the values that shaped them, and ensure that goals were aligned and not in conflict.

•**Toolkit format ideation**: This segment provided an opportunity for team members to explore potential formats for the toolkit, discussing which approaches would be most effective and practical.

•**Development of user 'proto-personas'**: Here, we began to define different user profiles to guide the toolkit's design, ensuring that it addressed the needs of its intended audience.

During the value elicitation part of the workshop, the AT presented its pro-justice approach to interpreting the AI Act. This included findings from a study on the landscape of AI ethics toolkits (as described above), highlighting how this research could inform the design of the new tool. For instance, the AT emphasised the importance of guiding teams through diverse stakeholder engagement activities, which, while not explicitly required by the Act, are essential for responsible AI deployment.

Among other, related ideas, the AT presented possible elements of the toolkit; for instance, guidance on including complaint mechanisms for groups affected by the use of AI, drawing on Sara Ahmed's suggestion that those who experience harms should not be made responsible for redressing them [1] (p. 24). Including a complaint mechanism, to use Ahmed's words, 'provides a lens, a way of seeing, noticing, attending to a problem in the effort to redress that problem' [1] (p. 24). In another example, the AT explored how feminist data science practices, such as those advanced by Catherine D'Ignazio and Lauren Klein, could inform sections of the toolkit focused on design and data management; while 'context is queen' in design, it is also crucial for 'conducting accurate, ethical [data] analysis' [10] (pp. 149, 91).

During the same session, the IT articulated their goal of developing a tool that could not only assist their own team in navigating the emerging European regulation, but also be made freely available to others, open access, so that any team working on high-risk AI tools could benefit. The IT emphasized the need for the toolkit to integrate socio-technical expertise to address the complex challenges arising from such systems.

The 'proto-persona' development exercise focused on defining the primary users of the toolkit. While the toolkit was designed to cater to a broad range of users, its primary target audience are product managers because they have a cross-functional oversight of AI projects in order to facilitate collaboration and deliberation both within AI development teams and across various organisational functions. We also wanted to enhance usability by making the toolkit flexible enough to support intra-organisational collaboration, allowing different compliance tasks to be assigned to the most appropriate people.

In conjunction with defining user personas, the teams discussed the pros and cons of different formats and agreed on a web-based application, which would enable flexibility and support iterative updates to reflect evolving approaches to both the AI Act and AI ethics more broadly. At the same time, it was agreed that the toolkit output should be made available as a downloadable PDF to serve as documentation of the process, including for sharing with internal and external stakeholders. Further reflections on the choice of format are discussed in section 4: Reflections, insights, and recommendations.

### 3.1.2 Prototyping and feedback response.
Prior to the second feedback session, a member of the AT did a thorough audit of every topic covered by, and every task implied by, the EU AI Act. Informed by this exhaustive list, the AT came together in online whiteboard sessions over several days to arrange and build a clear, but comprehensive, information architecture for the toolkit content.

The second collaborative working session then took place following initial feedback from Advisory Board members. During the session, both teams engaged in a collaborative activity based on rapid prototyping processes [2, 31]. This resulted in an initial interface layout and interaction wireframes, indicating functional features for the toolkit design template. Following a bottom-up prototyping process, the information architecture of the website to house the toolkit was defined through a navigation tree and first draft prototype using paper-based tools. Subsequently, the team developed a digital version using the collaborative tool Figma, allowing for an evaluation of the toolkit's usability prior to full-scale development. This collaborative exchange was instrumental in fine-tuning the toolkit's practical application while preserving its alignment with the pro-justice framework established in earlier phases. Following the session, the AT focused on developing content for each task, while the IT began constructing a high-fidelity prototype of the web application. This prototype would subsequently be shared with the Advisory Board and tested with users to gather additional feedback for refinement.

## 3.2 The composition of and engagement with a cross-sectoral, interdisciplinary board of advisors

### 3.2.1 Forming the Advisory Board.
To form a skilled multidisciplinary advisory board, we made a list of key areas of expertise critical to AI Act compliance, including, for example, law, policy, and computer science, and expended it to include experts in feminist and queer data science, disability justice, and critical design (see Table 1). This expertise was drawn from across academia, policy, and industry. The experts were essential to critiquing and improving our theoretical approach and offering advice for improved implementation.

We drew on a wide, combined network and recommendations from colleagues to invite candidates to be Advisory Board members. We invited feedback from members in two batches so as to reduce burden in our two-tier feedback process. The initial phase involved 12 members, while the subsequent round was expanded to include 23.

**Table 1: - Toolkit Advisory Board: Mapping expertise, Roles and contributions**

| **Expertise** (Disciplines and fields of expertise) | **Expert Roles** (Roles of contributing experts on the project advisory board) | **Types of Contribution** (Ways experts in this area contributed) |
| --- | --- | --- |
| Law | Lawyer, legal scholar | Compliance, rights |
| Technical | Computer scientist, software engineer, machine learning specialist, AI developer | Technical accuracy, needs of development process |
| Human-computer Interaction | HCI researcher, UX specialist, interaction designer | Considerations of user-facing tasks |
| Disability Justice | HCI researcher, disability advocate | Advice on tool accessibility; advice on interpreting access-related components of the act |
| Queer/feminist Data science | Feminist scholar, critical data scientist | Advice on inclusive data practices |
| Business | Product manager | Advice on feasibility and ease of use |
| Critical Design | Critical design researcher | Advice on alternative and critical design approaches |
| Critical Theory | Philosopher specialising in critical theory and its applications to AI | Provided theoretical foundations to process and the content of the tool |
| EU AI Act expertise | Policy advisor | Guidance on regulatory compliance |

*3.2.2 Soliciting feedback from the Advisory Board: round one.* During the first consultation, we disseminated two key working documents: the 'Explainer' and the 'Overview.' The 'Explainer' document introduced the toolkit, providing essential context regarding its intended purpose, target audience and pro-justice perspective. The 'Overview' document presented a skeletal outline of the toolkit's overall workflow with details on the planned content for each task.

Primarily, we requested feedback from advisors on how the proposed workflows aligned with, and addressed, obligations for high-risk AI systems and their providers. We also sought guidance on enriching the proposed tasks through the lens of feminist, anti-racist, and disability theory. Advisors also provided suggestions for additional tasks, resources, and real-world examples. They also helped to improve the tool's flow, with a particular focus on encouraging iterative and collaborative work processes during the compliance process.

*3.2.3 Soliciting feedback from the Advisory Board: round two.* For round two, we provided the Board with a web-based prototype, which included a complete draft of content for each task in the toolkit and implemented their feedback from round 1. The second round of the feedback process resulted in 177 comments spanning all areas of the toolkit, along with several general suggestions. These comments were grouped by relevant toolkit topics (or 'spaces', a term we explain in section 4) and each was carefully reviewed and addressed by the team. We categorised each piece of feedback as either within the scope of the toolkit, not actionable, or earmarked for future planning, acknowledging the limitations of our resources. The feedback process also helped us ensure that the toolkit comprehensively addresses all obligations outlined by the EU AI Act, with particular emphasis on high-risk AI systems and bearing in mind that the legislation was still evolving. We recognised the complexity of this task, given the Act's recent coming into force on 1 August 2024.

The board also helped to improve language and tone, identify areas where content could be reworded for greater clarity, and for making the language more engaging or thought-provoking. The goal was to create a resource that not only guides users through compliance but also stimulates critical reflection on the broader societal implications of AI development.

Usability was another key focus of our consultation. We asked the Board to evaluate the toolkit's interface, navigation, and overall user experience. This assessment included gauging the intuitiveness of the toolkit, identifying any tasks that might require further clarification, and suggesting improvements to enhance the overall flow and functionality. These insights were crucial in our ongoing efforts to create a tool that can effectively serve a diverse range of users.

## 3.3 User studies

In August and September 2024, we organised three user study sessions with three different prospective user groups with the aim of assessing the functionality of the tool and how it interacts; the activities were organised to test the toolkit's usability, acceptability, and feasibility within real-world AI development workflows. The participating groups were:

- The executive team of an SME AI provider (3 members);
- Responsible AI leads, AI developers, data protection officers, and the legal team from a multinational telecommunications company (17 members);
- A diverse group of ethical AI practitioners from various industries and disciplines, all part of a practice-oriented master's program (54 members).

Each session comprised three parts:

- **Introduction**: A brief overview of the AI Act's high-risk obligations and the toolkit's key features.
- **Initial evaluation**: Participants conducted a preliminary evaluation of the toolkit's overall navigation and content,

focusing on how easily they could understand and interact with the tool.

- **Detailed analysis and critique**: A thorough review of each 'space' in the toolkit (described in the following section) and how they interconnected with one another.

We asked participants to reflect on a series of questions to gauge the toolkit's effectiveness and integration potential, such as how easy it was to navigate and understand the toolkit, how effectively its structure supported ongoing, iterative processes of compliance, and which roles within their organisation might engage with the toolkit. The results of the studies and the discussion of how they affected our design choices are included in section 5.

Feedback collection activities on the functionality and content were not limited to the pre-release phase; in fact, the released version of the current toolkit is already in use and includes a mechanism for collecting feedback from users that will allow the team to improve its usability over time.

## 4 TOOLKIT DESIGN

Before we present reflections stemming from the process of designing the toolkit, we will briefly sketch the final structure and workflow of the toolkit itself. The free and open access toolkit is available at: https://aiact.cloud.ammagamma.com. A PDF version of the toolkit content is also available to download from the website homepage without the need to register.

### 4.1 Content and information architecture

The toolkits navigation logic is centred around the idea of topic 'spaces': interconnected areas of focus that team members engage with and revisit throughout software development and deployment. The toolkit identifies seven such spaces:

1. **Business, Team, Principles**: this space encourages teams to reflect on their composition, the core values driving their work, a critical analysis of the problem they aim to solve, as well as provide a justification for using AI.

2. **Impact, Risk, Mitigation**: teams assess potential harms associated with the system they are developing and devise strategies for managing risks.

3. **Stakeholder Engagement**: this space supports teams in consulting and involving various stakeholders during the design process.

4. **User-Centred Design**: teams focus on user-facing elements, ensuring accessibility, transparency, and meaningful consent, while also addressing principles like disability inclusion and design justice. Mechanisms for users to seek redress in case of harm are also emphasised.

5. **Data Governance**: teams document their methods for data collection and processing, ensure GDPR compliance, and explore principles of data justice.

6. **Model Governance**: this space provides guidance on documenting model training processes and selecting appropriate fairness metrics.

7. **Evaluation and Care**: teams are advised to monitor systems post-deployment, establish effective procedures for addressing reported issues, and manage complaints.

Each space begins with an introductory overview, followed by a series of tasks. These tasks provide structured guidance with consistent headings, including:

1. **Description**: explaining the task and its relation to the AI Act;

2. **Rationale** (and link to the Act): a clear explanation of why the task is worth doing and a link to where it appears in the Act;

3. **How to Approach**: offering step-by-step guidance for task execution;

4. **Expertise and Engagement**: identifying key stakeholders involved in the task; and

5. **Go Further**: offering optional, in-depth content to foster reflective thinking.

Each task is designed to scaffold the requirements of the AI Act, with a particular focus on justice-oriented and iterative ethical practices. Including a meaningful rationale for every task in the workflow supports autonomy and motivation [22] ensuring product managers understand exactly why that task is worth their time. The content subdivision into different tasks turns out to be necessary to separate which topics are required by law and which are the in-depth contents users can freely decide to include in the design process. The toolkit does not force the end-user to adopt a pro-justice approach; it is meant to be a tool for reflection to foster a more systemic and complex vision of AI systems, considering more heterogeneous contributions and points of view. To encourage active engagement, most tasks include a workbook section that requires user input to document and reflect on how they fulfilled the tasks within their organisations. This input can then be consolidated and downloaded as documentation for compliance or as preparation for future audits. This ensures the toolkit functions in three ways: as an educational resource for ethical AI, as a compliance tool that generates a documented record, and as a mechanism for continuous reflection and improvement.

### 4.2 User experience and features

Once logged in, the toolkit user is first presented with the 'Projects Page' where they can create new projects or click into existing ones. For each project, additional collaborators can be added and granted access from this page. Upon starting a new project, users can navigate its structure (see Figure 1) by following a non-linear process to complete the required tasks. Each task includes a series of instructions, as noted, including a workbook section – an editable form – that can be filled in to create what will become an exported PDF for documentation. Users can download a PDF version of the project containing all the completed information, which can be then used for compliance reporting purposes. All these functionalities are accessible through a user interface that has been designed according to usability and design principles that align with our theoretical approach.

## 5 REFLECTIONS, INSIGHTS, AND RECOMMENDATIONS

The final version of the toolkit – the result of the ideation, design, and testing process described above – was both shaped by and generated a number of insights into: translating AI ethics ideals into practice; deploying AI ethics expertise in real-life development contexts; and the toolkit format as a means of enabling compliance and helping teams move beyond compliance to engage with the latest AI ethics standards in their work. We present these insights in the following section under three broader categories: 1) AI ethics and compliance in practice; 2) toolkit design; 3) cross-sectoral collaboration.

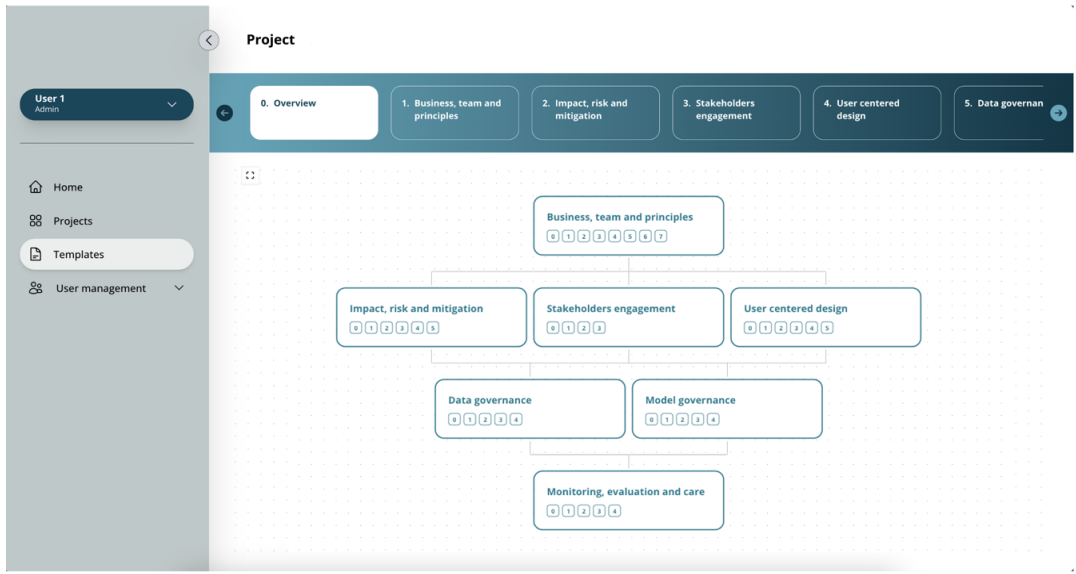

**Figure 1: Interface example, with navigation menu and spaces layout**

## 5.1 AI ethics and compliance in practice

*5.1.1 From box-ticking to ongoing, systemic evaluation.* From the outset of the project, the collaboration between the IT and AT teams was grounded in a shared commitment to compliance with the AI Act, underpinned by a pro-justice lens that aims to go beyond the bare minimum and concretise the Act's ambiguous elements. As our Advisory Board members advised, core to this aim is ensuring that the toolkit's users consider the broader social context of their (prospective) AI systems [6]. Getting to grips with the reciprocal and dynamic relationship between society and AI means recognising how societal forces shape AI development and, in turn, how AI influences and transforms social structures [9, 27]. Teams building AI products must understand that this interplay is not static but rather a continually evolving process that demands regular examination and reassessment. This is because AI systems are deployed in real-world contexts and their impacts emerge and evolve, often in ways that may not have been fully anticipated or understood during the initial impact assessment and development phase [28]. Consequently, as per our Advisory Board's suggestion, ethics toolkits must be designed to support continuous reflection and improvement in the development and deployment of AI systems. This feedback was in line with our initial background research on AI ethics toolkits and guidelines, which also consistently pointed to the need for tools and resources that abandon the 'box-ticking' logic that so many available resources engender, as we discuss in the Background section.

To address this need, we moved away from the initial workflow format that imagines development as a series of self-contained 'phases' – sets of tasks to be completed in a predetermined sequence – and into 'spaces' – interconnected areas of concern that you need to not only pass through but continue coming back to throughout AI development and deployment. These spaces are designed to be revisited throughout the development process. We achieve this by explicitly linking each space to the others where relevant, ensuring that users recognise the interconnected nature of the toolkit and understand the ongoing nature of responsible AI development. These adjustments aim to foster a more integrated approach that supports continuous reflection and iterative improvements. For example, in Space 5, focused on data governance, users are prompted to revisit Space 2 to reflect on data collection and processing impacts. They are asked to consider which data is collected, whose data is underrepresented, who has access to and analyses the data, and update their understanding of the potential negative impacts previously identified in Space 2. Additionally, we recommend that sections of toolkits dedicated to assessing the impact and risks associated with a given product and devising appropriate mitigation strategies (as we have done in Space 2) should be considered alongside content that supports engagement with affected stakeholders, as we have in Space 3.

Aside from this interconnected approach, we also take it as best practice to include advice on assessing and mitigating the environmental impact of AI systems and stakeholder engagement, even though neither are prescribed as mandatory by the Act. Despite this omission, compliance projects with a genuine commitment to AI ethics must consider the climate crisis and those most affected by negative effects of AI systems. To do this in a meaningful way,

Spaces 2 and 3 incentivise companies to take both seriously by providing practical methods and examples of how addressing these elements can be beneficial from a business perspective. Pro-justice is about thinking long-term about the value of ethical AI practices for both organisational success and societal impact, beyond the political roadblocks that prevent important issues being appropriately addressed by regulation.

### 5.1.2 Considering the ethics of business decision-making alongside design decision-making.
As we discussed in the Background section, our preliminary research pointed to the need for AI ethics resources that not only help teams consider the consequences of their design decisions but also facilitate an interrogation and adjustment of business objectives driving innovation. 'Space 1 – Business, Team, and Principles', is one of the toolkit elements designed to address this need and allow for a pro-justice approach in practice. Within Space 1, consecutive tasks guide teams through reflection on their own composition, helping users understand the importance of having an inclusive and diverse team, while cautioning against tokenisation. Space 1 also allows teams to analyse the core values driving their work, and how these relate to the EU's fundamental rights and values (and what they mean in practice). It also facilitates a critical analysis of the problem the team is trying to solve and the process of justifying the use of AI in a given area of application, with a reference to existing AI incidents databases.

Crucially, and in response to the Advisory Board's feedback, Space 1 also points to the AI 'exit-door' strategy: we make clear the decision not to use AI to solve a specific issue must also be considered viable. This decision may result from the analysis of the problem the team is trying to solve, as its members review previous AI incidents. Further, also responding to the Board's feedback, users are guided to assess labour conditions throughout the AI lifecycle (including within their own team) and to actively question how roles are distributed along lines of gender and race. This helps ensure that both the AI systems and the development processes do not perpetuate or reinforce social stereotypes and inequalities.

### 5.1.3 Defining AI ethics 'expertise' as collective.
For the purposes of this project, we had to decide what qualifies as AI ethics expertise. As we have outlined in 2.3, we view expertise as a collective endeavour that arises through conversations between people with different kinds of insight into AI systems. Our Advisory Board reflects this approach (see table 1). The process of gathering expertise involved conversations between our team and the Board when discussing and processing their feedback. Our tool attempts to retain the complexity of the issues raised and in as useful a way as possible by demonstrating a variety of practical approaches, as well as the negotiations and compromises ethics necessarily entails. Our tool is therefore a repository of expertise, which is reflective of our commitment to recognising that AI ethics work is not a matter for a single consultant but is distributed among many actors.

## 5.2 Toolkit design

### 5.2.1 Balancing a non-linear evolving reality with the need for achievable objectives.
As mentioned above, the team's value commitments included resisting the notion that the pursuit of responsible AI is a linear process that can be quantified, completed, and checked off.

Such a perspective risks oversimplification and reductionism [26], undermining the nuanced and complex nature of ethical decision-making. As such, we began to explore representing user tasks in a circular and iterative manner. However, if the toolkit had no way to track progress and felt impossible to complete, it could be overwhelmingly daunting.

During the second two-day session, the team identified this tension between a circular approach and the comparatively linear lifecycle models commonly applied for AI systems, which follow distinct phases such as ideation, development, and deployment. Even where iteration is acknowledged within development (for example, in agile environments), phases and projects still have clear end points. This allows for sufficient certainty when planning against personnel and resource constraints. The IT noted that balancing the need for ongoing reflection and improvement with the time-sensitive nature of industry workflows presents practical challenges. However, they also recognized the value of facilitating an iterative approach in fostering long-term ethical commitment, driving changes to organisational culture, and ensuring compliance with evolving regulations, particularly in high-risk AI scenarios.

As a result, we revised our initial plan to exclude any means of marking task completion within the toolkit interface and aimed instead to add visual cues that could track progress without closing off sections and possibilities. Moreover, the precise design was not finalised during the collaborative working session. We agreed that user testing would be essential in helping us determine the most effective way to present this progress. Later in the user testing session, we sought design input from participants on how to provide users with a sense of progress without oversimplifying tasks into discrete, one-time actions. In the end, we decided to use visual cues, for example using colour rather than checkmarks and presenting the menu as an ordered and interconnected but non-linear path. This conveys progression without reducing it to quantifiable percentages.

Finding ways to soften signifiers of completion and to convey non-linearity, dynamism and longer-term stewardship, shows how interface design becomes part of the AI ethics discourse. The tools we create shape mental models, and these can perpetuate misconceptions or improve them, thus shaping practice. Such an approach brings to life reflections from scholars including Jasanoff, who advocates for the development of 'technologies of humility' that encourage reflection on ambiguity, indeterminacy, and complexity, moving beyond the use of a binary logic especially in the presence of reasonable disagreement [18].

### 5.2.2 Balancing inherent ambiguity with the need for actionability.
Unpacking the often vague specifications of the Act, while enhancing it with rich insights from critical research, creates an education challenge that necessarily involves engaging with open-ended concepts, contested ideas, and evolving methods. These must still, however, ultimately lead to specific, documentable action for the users of the toolkit. If we had chosen a report format, rather than an interactive software system, we might have leaned too heavily into theoretical explanation and provided little help for practitioners looking for specific guidance on what to do and how to do it. An interactive format allows us to create the toolkit composed entirely of 'Tasks' rather than topic sections to prevent this. The other key approach to guaranteeing every section balanced education with action was the inclusion of the workbook. Tasks themselves might have been vague had we not needed to include specific instructions with text input fields within each one, ensuring their specificity and documentability.

Advisory Board members provided valuable suggestions for incorporating more actionable items within workbook sections. These included: adding more input fields so people could document with more granularity; making use of variable terms like 'transparency' more consistently (in this case, we used only the AI Act's definition which is distinctive); consolidating content for greater simplicity; and ensuring that those sections of the workbook intended to be used multiple times throughout the development process (e.g. 'potential misuse' as further use-cases arise) were clearly identified as such. This feedback played a critical role in transforming the toolkit into not just an information resource but a practical tool that guides users through concrete steps linked to documentation for their compliance process.

User testing feedback indicated that the navigation of the toolkit was intuitive and comparable to product management applications. However, participants suggested improvements for better integration into existing workflows, such as offering the toolkit as a plugin for common project management software. User testing also revealed the importance of establishing clarity around the toolkit's role in supporting compliance with the Act. Several participants raised questions such as, 'If I complete all the tasks within the toolkit, does that mean I'm fully compliant?' Therefore, as part of the onboarding process, we introduced a 'What you can expect?' section to manage expectations accordingly.

*5.2.3 Balancing generality and standardisation with specific industry sector needs.* Although the AI Act is designed to be applied across industries (as is the toolkit) the reality is that different sectors and businesses will have different needs with respect to AI workflows. For example, high-risk solutions concerning critical infrastructure (e.g. safety components in energy supply) and AI systems used in workforce management (e.g. recruitment tools), will have a different focus with respect to explainability of the model. Different values and priorities will guide responsible AI decisions across companies. Moreover, companies develop internal sets of best practices which, in some cases, could be even *more* strict than those described in the Act. Finally, different projects will be at different stages of responsible AI development; some will be retrofitting while others will be starting from scratch.

The IT highlighted this potential mismatch between a 'one-size-fits-all' approach to a toolkit and the realities of diverse industry practice. This pointed to a need for the resource to be customizable to make it fit more seamlessly into organisational workflows and respond better to company needs. We approached this legally, through our choice of content license, and technically, through the integration of templating. Specifically, we made the toolkit content available under a Creative Commons licence which allows anyone to freely use and customise the content to fit their needs. In this way, companies can add content, remove sections that do not pertain to them, and supply pre-filled answers to questions that are common across projects (e.g. company values). Moreover, to facilitate this practically, the IT added a templating system that allows users to create a copy of the default toolkit and then edit it directly. This allows for the future creation of organisation-specific versions of the toolkit to meet diverse needs while remaining aligned with the Act.

## 5.3 Cross-sectoral collaboration

*5.3.1 Mitigating the impact of shifting institutional priorities.* At the beginning of the second year of the collaboration, the IT underwent significant organisational changes following its acquisition by a global technology consultancy firm. Such acquisitions are not uncommon in the AI sector, as acquiring AI-based businesses has become a critical business strategy element for enterprises seeking to position themselves as market leaders [12]. The institutional shift for the IT introduced a range of challenges to our cross-sectoral collaboration related to team restructuring, shifts in time availability, and a reassessment of the project's macro-objectives. For example, we urgently had to consider how the toolkit related to the consultancy firm's global strategy to develop a comprehensive Responsible AI Platform. Critical questions emerged: *How could the toolkit integrate with this broader vision? Would offering the toolkit freely, alongside tailored offerings for other tools, remain a viable strategy?*

Leadership transitions and team restructurings are common challenges in long-term, collaborative projects; in academia-industry partnerships in particular such changes may require additional discussion of project objectives as institutional priorities change. In our case, the leadership transition and restructuring on the IT end did not become a cause for concern. In the revised institutional framework, the toolkit evolved into a strategic asset within the acquiring company's larger Responsible AI Platform, which it offers globally, and which the AT was happy with. This is in part because the framework ensures the pro-justice toolkit's impact is a tangible, openly accessible resource for institutions and organisations of varying scales, as well as a tool of responsible AI education within the acquiring company itself. At the same time, we acknowledge that this shift – the arising need for further value re-alignment between teams and the re-prioritisation of action points on a new project timeline – could have had a negative impact on the overall outcome, had the project aims not matched the acquiring company's objectives. A lesson learned for other teams working in academia-industry partnerships and for our own future work is that such potential shifts should be discussed and mitigation strategies considered before the collaboration begins, as early as possible in the partnership agreement negotiation process.

*5.3.2 Ensuring ongoing toolkit maintenance and longevity.* Given the evolving nature of the AI Act and the associated regulatory requirements, it became essential to establish a robust strategy for maintaining the toolkit and ensuring its long-term viability. As the regulatory landscape is expected to mature over time, periodic evaluation will be necessary to ensure that it remains aligned with updated requirements. Without alternative ongoing funding for maintenance, a key factor in ensuring the online toolkit's sustainability is securing a return on investment for the consultancy firm maintaining the interactive application while ensuring that the resource remains open access and free to use. Leadership changes, if coupled with a lack of measurable benefits, could risk deprioritisation or abandonment of the toolkit. To mitigate this, the consultancy firm has devised a monetisation strategy that allows its own clients to instantiate customised copies of the toolkit on private cloud infrastructures, while the main version of the toolkit remains open access and freely accessible. This approach enables continuous updates aligned with the evolution of the AI Act and iterative development of the toolkit, securing its long-term relevance and effectiveness.

## 6  FUTURE RESEARCH AND CONCLUSION

Herein we have described the process, challenges, and lessons learned from a one-year cross-sector partnership developing a projustice software-based toolkit to guide developers of high-risk AI systems through compliance with the EU AI Act. The final toolkit is designed to promote continuous engagement with the ethical dimensions of AI work, reflection and iteration rather than boxticking while attending to the pragmatic needs of industry development practice.

While the toolkit targets the needs of product managers and their teams, the other essential half of the equation is citizens (who either act as the end-users of AI systems or are subject to AI decision-making). For example, just as organisations must provide mechanisms for feedback and redress, citizens must understand these exist and how to access them. By the same token that companies must understand their responsibilities, citizens must understand what protections they can expect and cannot expect from AI technologies. Therefore, our future work will explore the needs of citizens and how best to support them in exercising their rights and responsibilities within the AI ecosystem.

Real use case testing is another priority for future work. Testing the effectiveness of the toolkit and adjusting it accordingly requires the pursuit of strategic collaborations with appropriate institutions. Our first step will be to present the toolkit to institutional networks, including the Digital Services Office of the European Commission, and proposing its evaluation within a European Digital Innovation Hub. Such initiatives not only foster institutional adoption but also position the toolkit as a candidate for inclusion in regulatory sandboxes established by EU member states to explore and refine AI compliance frameworks.

There are also opportunities for research on the needs of low to medium risk technologies, as well as domain-specific guidance. We hope that by sharing the challenges and insights from our journey, future teams developing tools for responsible AI can draw on these experiences to shape their own roadmap and anticipate potential obstacles. While the Act represents a crucial step toward beneficial AI, there remains significant work to connect its directives with industry practices. With this toolkit, we aim to take a step in that direction.

## Acknowledgments

Funding for this project has been provided by a grant from Stiftung Mercator, grant number 200446.